\documentclass[showpacs,twocolumn,floatfix,aps, pre
]{revtex4-1}
\usepackage{epsfig}
\usepackage{amsmath}
\usepackage{subfigure}
\usepackage[colorinlistoftodos]{todonotes}
\usepackage{enumitem}
\usepackage{tikz}
\usetikzlibrary{shapes,arrows,shadows}
\usepackage{bm}
\usepackage{color,soul}
   
\newcommand{\beq}{\begin{equation}}  
\newcommand{\eeq}{\end{equation}}  
\newcommand{\bea}{\begin{eqnarray}}  
\newcommand{\eea}{\end{eqnarray}}  

\begin{document}

\title{Grain Scale Modeling of Arbitrary Fluid Saturation in Random Packings}

\author{Konstantin Melnikov$^1$}

\author{Roman Mani$^{1}$}

\author{Falk K. Wittel$^{1}$}

\author{Marcel Thielmann$^{1}$}

\author{Hans J. Herrmann$^{1}$}
\affiliation{$^{1}$Computational Physics for Engineering Materials, ETH Z\"urich, 
Stefano-Franscini-Platz 3, 8093 Z\"urich, Switzerland}

\date{\today}

\begin{abstract}
We propose a model for increasing liquid saturation in a granular packing which can account  for liquid redistribution at saturation levels beyond the well-studied capillary bridge regime. The model is capable of resolving and combining capillary bridges, menisci and fully saturated pores to form local liquid clusters of any shape. They can exchange volume due to the local Laplace pressure gradient via a liquid film on the surfaces of grains. Local instabilities like Haines jumps trigger the discontinuous evolution of the liquid front. The applicability of the model is demonstrated and compared to benchmark experiments on the level of individual liquid structures as well as on larger systems. 
\end{abstract}


\maketitle

\section{Introduction}
The study of liquid distribution and transport in unsaturated porous media, such as granular packings, has long been a topic of intense cross-disciplinary research. The mechanical properties of granular materials are significantly affected by their liquid saturation \cite{Herminghaus2005, Mitarai2006} as liquid clusters emerge and grow in size. While the liquid bridge regime at low saturation is well-studied numerically and experimentally \cite{Richefeu2006, Scholtes2009, Mani2012, Mani2012b}, the extension of models to higher liquid contents remains a demanding task. A better understanding of this regime is however crucial for solving a number of open problems in science and engineering including e.g. rainfall-induced slope failures, oil recovery or flow in porous media, just to name a few.

Early works in this field employed the so-called ideal soil model, consisting of uniform solid spheres in a regular packing \cite{Haines1927, Haines1930}. In the 1950s, network models emerged, originally proposed by Fatt \cite{Fatt1956}. They represent only the pore space by sites (pore bodies) of arbitrary shape and position, interconnected by bonds (pore throats) to form a network \cite{Blunt1992, Fischer1999, Blunt2001, Lowry1995}. Network topologies are either obtained from thin section analysis \cite{Vogel2001}, tomographic measurements \cite{Coles1998}, from fundamental statistical assumptions \cite{Lowry1995} or directly from packed particle configurations \cite{MellorPHD}. Even though the geometrical representation of the pore space is significantly simplified, important information can be obtained for two-phase flows such as relative permeability \cite{Fischer1999}, trapped immiscible fluid \cite{Lowry1995},  drainage and imbibition \cite{Mason1995}, or capillary pressure - saturation relations \cite{Held2001}. A major limitation of pore network models lies in the static representation of the pore space with respect to deformations, where network parameters would constantly change.

To cope with such challenges, a more geometrically detailed representation of the pore space and of the fluid interfaces is needed. While the pore network is still derived from a random sphere packing that could evolve in time, details on liquid interfaces like liquid bridges and menisci of constant curvature \cite{Haines1927} are considered \cite{Gladkikh2005, GladkikhPHD, Bryant2013}. In these grain-based fluid invasion models of three dimensional porous media, the fluid front is driven by local instabilities and correctly reproduces drainage and imbibition experiments.

With the availability of advanced microtomography, a rich variety of liquid clusters was found \cite{ScheelPHD, Scheel2008nature}. The number and the size of observed liquid clusters was shown to strongly depend on the saturation level. Inspired by these findings, we propose a model that explicitly considers all possible liquid morphologies on the pore scale in a sphere packing. Our aim is to be able to simulate the entire range of saturation levels from the dry state, via cluster formation and growth to the fully saturated state. Even though we use a discrete element model (DEM) with spherical particles to simulate the random packing, we fix all particle positions for the fluid simulation. Hence we describe the first step on the way to a general model that combines two-phase flow and deformation. 

First, in chapter~\ref{sec:model} we give a comprehensive description of the model with respect to implemented liquid structures and their composition into larger clusters along with geometrical stability criteria for their growth and decay. In the following chapter~\ref{sec:codeStructure} we provide a brief overview on the simulation procedure. We show applications of our model on two validation experiments; one on the pore scale for the trimer formation and decay and the other one for the evolution of cluster distributions in large systems (chapter~\ref{sec:Validation}). Finally we show a typical simulation of the cluster evolution for the case of liquid injection at a singular point, before we summarize main results and draw conclusions (chapter~\ref{sec:Conclusion}). In Appendix~\ref{sec:regula_falsi} we describe in detail the important pressure update algorithm for the volume controlled simulations.
\section{Components of the Grain Scale Model} \label{sec:model}
Our aim is to represent both the pore space and the liquid structures as close to their real shape as possible, accepting drawbacks in numerical performance. Hence the pore-throat network is constructed based on the exact geometrical positions of a previously simulated granular packing. Liquid clusters are represented as a combination of elementary units such as liquid bridges, menisci or entirely filled pore bodies. Not only units can form higher geometrical configurations such as trimers, pentamers, tetrahedral clusters and higher ones, but they can also decompose into elementary ones. This evolution is determined by stability criteria discussed in this chapter as well.

\begin{figure}[htb]
\centering
\includegraphics[width=0.3\textwidth]{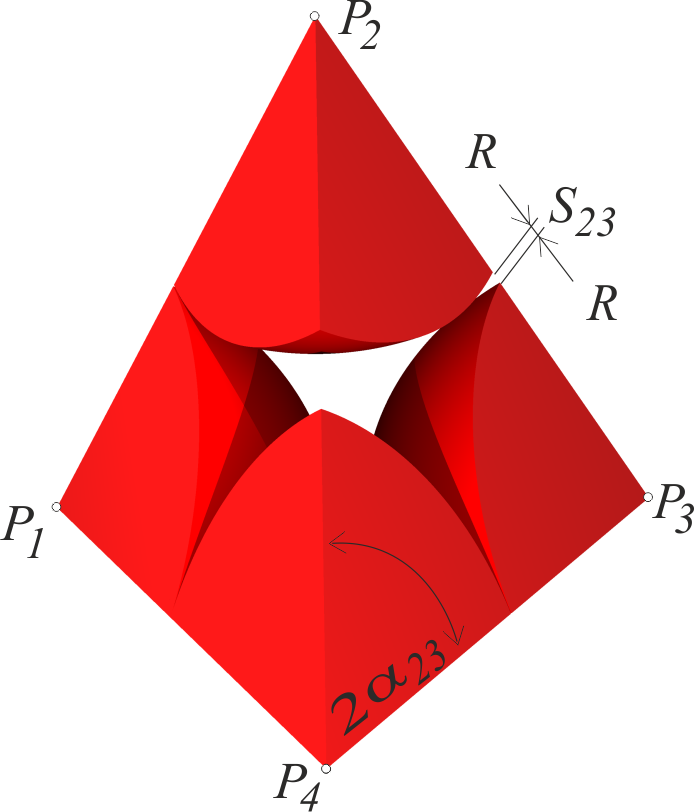}
\caption{\label{Fig:cell} (Color online) Single tetrahedral cell from a triangulated particle packing on sphere centers $P_{i},$ $i=1,..,4$ with radius $R$. $S_{23}$ denotes the contact distance of particle 2 with respect to particle 3 and $\alpha_{23}$ the respective opening angle.} 
\end{figure}  
We extract the pore space from the initial particle packing given by the positions of the sphere centers and their radius $R$ via Delaunay triangulation following Ref.~\cite{MellorPHD}. Hence the entire sample volume is subdivided into single tetrahedra like the one shown in Fig. \ref{Fig:cell}. The void space in each tetrahedron is called the pore body while the cutting areas of the pore body with the respective faces of the tetrahedron form the four pore throats of a cell. This way, a pore network is extracted directly from the topology of the sphere packing. Pore bodies can be empty, partially filled with liquid separated by menisci or entirely filled by liquid. In the discussed model we assume that the pressure inside the liquid phase is smaller than the gas pressure.
\subsection{Representation of Liquid Structures}\label{sec:structures}
\begin{figure*}
\centering
\includegraphics[width=1\textwidth]{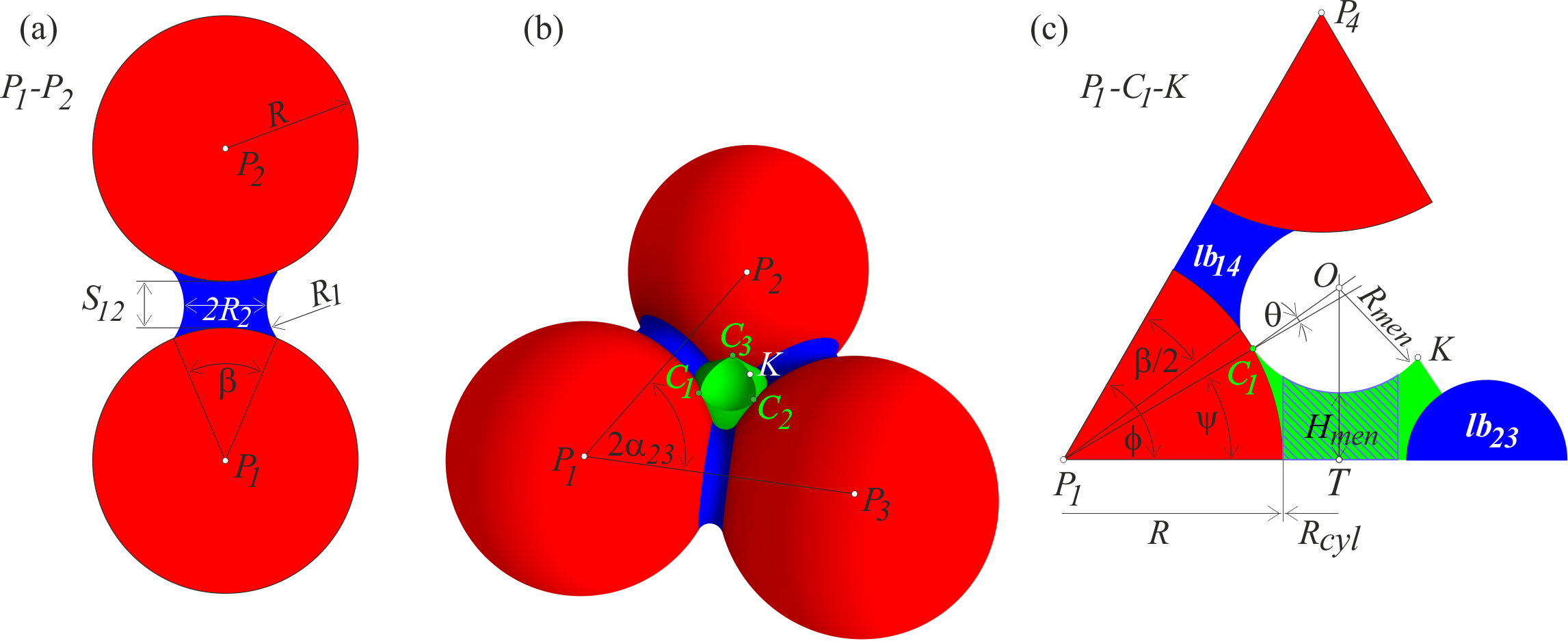}
\caption{(Color online) (a) Inter particle liquid bridge with separation distance $S_{12}$ and filling angle $\beta$ shown in the cross section through the centers of the particles $P_{1}-P_{2}$. $R_{1,2}$ denote the principal radii of the curvature. (b) Trimer: Meniscus in the pore throat which is build by three grains with centers $P_{1,2,3}$. The concave meniscus (green) has one contact point ($C_{1,2,3}$) with  each neighboring particle. Additionally, connected liquid bridges (blue) are shown. (c) Enlarged cut through $P_1$-$C_1$-$K$ showing the meniscus position in the pore throat. The point $K$ is located between $C_{2}$ and $C_{3}$ on the meniscus, see (b). The point $O$ is the center of the meniscus. $lb_{14}$ and $lb_{23}$ denote the liquid bridges between the particles $P_{1}$ and $P_{4}$ respectively $P_{2}$ and $P_{3}$. The hatched area shows the cut through the cylinder which approximates the meniscus volume. Figure (c) is based on \cite{GladkikhPHD}.}
\label{SphericalCap}
\end{figure*}

When \textbf{pore bodies} are filled with liquid, they are considered as one of the three building units of the liquid clusters, while liquid bridges and menisci shown in Fig.~\ref{SphericalCap} are the other two. A \textbf{liquid bridge} is located between two grains (Fig.~\ref{SphericalCap}(a)). The pressure difference between gas and liquid phase $\Delta P$ = $P_{liquid} - P_{gas}$ is due to surface tension $\gamma$ and well described by the Young-Laplace equation \cite{deGennes2003}
\begin{equation}
   \Delta P=\gamma C = \gamma \left(\frac{1}{R_{1}} + \frac{1}{R_{2}}\right),
\label{LaplaceEquation}
\end{equation}
where the surface curvature $C = 1/R_{1} + 1/R_{2}$ is defined by the principal radii of curvature $R_{1}$ and $R_{2}$. This equation can be solved numerically for a liquid bridge with boundary conditions defined by the contact angle $\Theta$. Since solving the Young-Laplace equation is a numerically expensive and difficult task, we interpolate the capillary pressure $P$ and the volume $V$ of a liquid bridge from tabulated values. These  were calculated by Semprebon \textit{et al.}~\cite{Mani2015} using the numerical energy minimization method of the software Surface Evolver \cite{Brakke1996}. Both pressure $P$ and volume $V$ are functions of the separation distance $S$ between the grains, the contact angle $\theta$ for solid-liquid interfaces and the filling angle $\beta$ of the liquid bridge. Note that $\theta$ is kept constant at $5^{\circ}$ throughout this work.

A key parameter for a liquid bridge and its pressure is the separation distance $S_{ij}$ between the grains $i$ and $j$. Note that this distance remains constant during the fluid simulation, however capillary bridges can rupture due to liquid outflow. The dimensionless rupture distance $S_{c}$ of the liquid bridge (in units of the particle radius $R$) is related to the dimensionless volume $V$ and the contact angle $\Theta$ through the empirical expression derived by Willett \textit{et al.}~\cite{Willett2000}:
\begin{equation}
S_{c} \simeq (1+\tfrac{1}{2}\Theta)(\sqrt[3]{V}+\sqrt[3]{V^2}/10).
\label{equ:ruptureDist}
\end{equation}

An important assumption of our model, motivated by ideas of Haines \cite{Haines1927}, is that the liquid-air interface between three grains, called \textbf{meniscus} is of spherical shape with constant curvature (Fig.~\ref{SphericalCap}). A cross section perpendicular to the pore throat is shown in Fig.~\ref{SphericalCap}(c). This approximation is in satisfying agreement with experimental observations \cite{ScheelPHD, Scheel2008nature}. The centers of the four possible menisci inside a tetrahedral cell are located on the normal of each pore throat through the circumcenter of the respective face. The exact position on this normal can be calculated once the contact angle with the grains $\Theta$ and the meniscus radius $R_{men}$ are known, following the method proposed by Gladkikh \cite{GladkikhPHD} (see Fig.~\ref{SphericalCap}). The pressure drop between gas and liquid is again determined by the curvature of the interface $1/R_{men}$ via the Young-Laplace equation (Eq.~\ref{LaplaceEquation}). Due to the spherical shape of the meniscus the negative pressure drop is given by
  \begin{equation}
   \Delta P= \gamma \cdot \frac{2}{R_{men}}.
   \label{equation:meniscusPressure}
  \end{equation}
In reality single menisci cannot exist, but always occur in combination with associated liquid bridges forming one liquid body with equal Laplace pressure and common surface (see Fig.~\ref{SphericalCap}) \cite{ScheelPHD, Scheel2008nature, Scheel2008}. We adopt the assumption of equal Laplace pressure within the liquid body, determined by the curvature of the meniscus. Note that this pressure can fall below the values for stable bridges, resulting in a liquid body with less than three associated liquid bridges. In small clusters menisci with less than three associated liquid bridges have not been observed \cite{ScheelPHD}. Therefore, we allow for menisci with a reduced number of bridges when clusters have more than one filled pore body. This assumption is required to enable clusters to fill also those regions of the pore space where inter particle separations are large.
\begin{figure}[htb]
\centering
\includegraphics[width=0.45\textwidth]{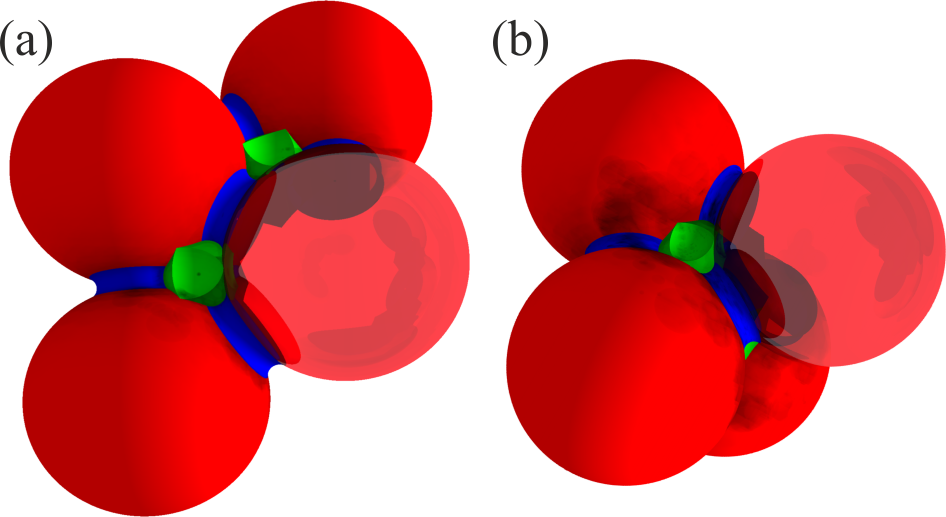}
\caption{(Color online) Small liquid clusters: (a) Pentamer and (b) tetrahedral cluster.}
\label{Fig:clusters}
\end{figure}
 
A main innovation of our approach is the possibility for the formation of local \textbf{liquid clusters} which can evolve inside the granular material. They are composed of the three basic units introduced above: filled pore body, liquid bridge, and meniscus. In principle every liquid body with the exception of a single liquid bridge can be considered as a cluster. The smallest possible cluster is called trimer (Fig.~\ref{SphericalCap}b) \cite{Scheel2008nature}. In our model it is bounded by two menisci that are located in neighboring tetrahedra on both sides of the common pore throat and their shared liquid bridges. Higher order clusters like pentamers are formed when two trimers share one liquid bridge (Fig.~\ref{Fig:clusters}a). The smallest cluster with a filled pore body is a tetrahedral cluster with the liquid body bounded by four menisci and six associated liquid bridges (Fig.~\ref{Fig:clusters}b). 

The control variable in our model is the volume, thus its correct calculation for the liquid clusters is essential for any simulation. Remember that inside a liquid cluster, pressure is homogeneous and given by the radius of its menisci $R_{men}$. In general, the volume of a cluster $V_{c}$ with $N_{imb}$ imbibed pore bodies and $N_{men}$ menisci can be written as
\begin{equation}
V_{c}(R_{men}) = \sum_{i=0}^{N_{imb}} V_{pore, i} + \sum_{j=0}^{N_{men}} V_{men, j}(R_{men}), 
\end{equation}
where $V_{pore, i}$ denotes the volume of the imbibed pore body $i$ and $V_{men, j}$ the volume of the meniscus $j$ including the associated liquid bridges of this meniscus. Note that liquid bridges in a cluster are always associated with one of its menisci creating a common liquid body. The volume of the filled pore is calculated by subtracting the partial volumes of the four particles contained within the tetrahedral cell from the volume of this cell. 

The volume beneath a meniscus is approximated by the volume of a cylindrical body $V_{cyl}$ coaxial to the normal of the pore throat, see Fig.~\ref{SphericalCap}(c). Its radius $R_{cyl}$ equals the circumradius $R_{circ}=P_{1} T$ minus the particle radius $R$, and its upper surface is bounded by the meniscus. Note that if the meniscus bounds a filled pore body, the meniscus can intersect the pore throat. In this case the volume of the pore body beneath the throat plane must be reduced by the volume bounded between the meniscus and the throat pore plane. As mentioned above, to calculate $V_{men, i}$ we also take into account volumes of the liquid bridges $V^{lb}_{i}$ associated with the meniscus $i$:
\begin{equation}
V_{men}(R_{men}) = V_{cyl}(R_{men}) + (0.5+\epsilon)  \sum_{i=0}^{m} V_{i}^{lb}(R_{men}).
\label{eq:menVol}
\end{equation}
The geometrical correction parameter $\epsilon$ accounts for the volume excess of a real meniscus with the connected liquid bridges, compared to our approximation in which $V_{cyl}$ underestimates the volume bounded by the meniscus (see Fig.~\ref{SphericalCap}(c)). Depending on the number of the connected liquid bridges the index $m$ can take values between 0 and 3. If a meniscus shares a liquid bridge with an adjacent meniscus (e.g. for a pentamer, Fig.~\ref{Fig:clusters}a) the number of connected liquid bridges for one of the menisci reduces by one. With this rule we assure that the bridge volumes are not counted twice. Note that for every meniscus only half of each liquid bridge volume is considered, since the other half is located in the opposite triangulation cell if its pore body is not saturated (see the liquid bridge $lb_{14}$ between the particles $P_{1}$ and $P_{4}$ in Fig.~\ref{SphericalCap}(c)).
\subsection{Evolution of Liquid Structures} \label{sec:stabilitCriteria}
Since we assume all particles to be fixed in space at this stage, liquid structures can evolve when their volume changes due to condensation or evaporation at gas-liquid interfaces or the injection of liquid at distinct positions. Inside of a liquid body the flux is instantaneous, while transport between bodies $i$ and $j$, sharing the same particle occurs through a liquid film on the particle surface. The relevance of liquid transport through this wetting layer was experimentally proven by Lukyanov \textit{et al.} \cite{Lukyanov2012}. Under the assumption of stationary flow in the film, the volume flux is considered to be proportional to the local capillary pressure gradient $\Delta P = P_i-P_j$. The dimensionless flux into a liquid structure $i$ is then given by the sum of all volume fluxes between $i$ and the connected structures $j$:
\begin{equation}\label{eq:liqu_trans}
\dot{V_{i}} = \frac{R}{\gamma}\cdot\sum_{j=0}^{N_i} \mu_{ij} (P_{j} - P_{i}),
\end{equation}
where $N_{i}$ is the number of all liquid structures connected to $i$. The dimensionless conductance coefficient $\mu_{ij}$ must include effects associated with the geometrical distance between structures, the number of structures connected to one particle etc \cite{Mani2015,ManiPHD}. However, for simplicity we assume this coefficient to be $\mu_{ij}=0.01$ for all calculations in this paper. The value of $\mu_{ij}$ defines the time scale of the pressure equilibration between connected liquid structures. Note that this liquid transport mechanism is in particular important at low saturation levels. 

The propagation of the liquid interface in the material is triggered by local instabilities. Hence we model interface evolution as a discontinuous process with instantaneous jumps from one stable configuration to the next one depending on the local volume, if the corresponding pore body is drained or imbibed. An initially stable interface can become unstable due to curvature changes, when the volume of the liquid body increases or decreases. We implemented five geometrical instability criteria similar to the ones described by Motealleh \textit{et al.} \cite{Bryant2013}. The first four result in growth while the last criterion accounts for drainage resulting in shrinkage of the liquid cluster:
  
\textbf{Criterion c1:} Pore throats are filled due to coalescence of liquid bridges if they touch each other, creating a new trimer. The criterion is based on the filling angles $\beta_{1}$ and $\beta_{2}$ of the bridges and the opening angle of the corresponding pore throat $2\alpha_{ij}$. If $0.5(\beta_1+\beta_2)>2\alpha_{ij}$ the pore throat is imbibed (see Fig.~\ref{SphericalCap}(b)).
   
\textbf{Criterion c2:} Imbibition of a pore body due to an interface instability is triggered when a meniscus touches a single liquid bridge inside the tetrahedral cell that was up to now not part of the liquid body (Melrose criterion, \cite{Melrose1965}). The transition from trimer to tetrahedral cluster (Fig.~\ref{Fig:clusters}a$\rightarrow$b) is such an example. Using the filling angle of the meniscus $\psi$, the configurations become unstable when $\psi+\beta/2>\phi$ with the face-edge angle $\phi$ shown in Fig.~\ref{SphericalCap}(c). The calculation is described in detail in Ref.~\cite{GladkikhPHD}.
   
\textbf{Criterion c3:} Imbibition of a pore body due to an interface instability can also be triggered by the merging of two menisci in the same pore body. The criterion is a generalization of the Haines imbibition criterion to non-zero contact angles. If two menisci centers reach the same point, the menisci touch and build a single spherical interface. Then the menisci become unstable, disappear and the pore body is filled \cite{Bryant2013}. Note that in situations where all four possible menisci of a tetrahedral cell exist, a gas bubble gets trapped. 
   
\textbf{Criterion c4:} When a meniscus touches the opposite particle, the pore body gets entirely filled \cite{Bryant2013}. This event occurs if the curvature of the meniscus $1/R_{men}$ is small enough to allow a touch of the meniscus and the opposite grain (the exact value depends on the geometry of the particular triangulation cell). Note that this criterion is an extension of the one proposed in Ref.~\cite{Cieplak1990} from two to three dimensions.
   
\textbf{Criterion c5:} Decreasing volumes result in increasing absolute values of Laplace pressure that can trigger two different kinds of drainage: \textbf{(c5-1)} the decay of trimer units of clusters and \textbf{(c5-2)} the drainage of pore bodies. Hence the first one describes breakup of a liquid body inside a single pore throat. Due to the assumption of a meniscus with constant curvature, an intuitive criterion of touching menisci on opposite sides of the pore throat overestimates the stability of the trimers by about 20\% with respect to experimental observations \cite{ScheelPHD}. We propose a criterion based on a minimum thickness for stability:
\begin{equation}
H_{men}^{min} \geq \kappa R,
\label{equ:drainageInst}
\end{equation}
with the adaptable drainage parameter $\kappa$ to calculate the critical height $H_{men}^{min}$ of menisci for stability. Best agreement with experiments is obtained for a value of $\kappa=0.15$. Depending on the opening angle $\alpha_{23}$ (see Fig.~\ref{Fig:trimerDecay}), two or three liquid bridges remain.
The drainage criterion \textbf{(c5-2)} is related to drainage of a single pore body. When the center of the meniscus of a neighboring cell touches the respective pore throat plane of an entirely saturated cell, its pore body is drained. Instantaneously the liquid interface jumps to a new stable position, creating three new menisci in the pore throats of the drained pore body, removing the responsible meniscus in the neighboring cell. This criterion follows the ideas in Ref.~\cite{Bryant2013} and is an extension of the original one proposed by Haines \cite{Haines1927} to arbitrary contact angles. Now that the model description is complete, we address its simulation.
\section{Numerical Implementation} \label{sec:codeStructure}
Prior to the fluid simulation, we construct the pore space, defined by a dense particle packing, using a discrete element method (DEM) described in detail in Ref.~\cite{ManiPHD} with contact dynamics. We perform a random sequential adsorption \cite{Feder1980} of equally sized spheres inside the sample volume $V_{sample}$ with periodic boundary conditions. To achieve a desired packing density $\rho$, particle radii $R$ are increased while particles are allowed to rearrange. A small amount of liquid is assigned to every particle to stay in the pendular regime where only liquid bridges exist. Whenever particles contact, a liquid bridge is created; when particles separate beyond $S_c$, bridges rupture (\ref{equ:ruptureDist}). In the following, the liquid content is defined using the total volume of the liquid structures $V_{liquid}$ as $W_{c} = V_{liquid} / V_{sample}$. Finally, particles are fixed at their positions for the remainder of the simulation, liquid bridges are allowed to equilibrate pressure, and the sample is subdivided into tetrahedral cells by a periodic Delaunay triangulation of particle centers using the CGAL package \cite{GGAL_Periodic3D}.

In our implementation the volume is the control variable. Liquid is either injected (removed) at distinct point in space or condensed (evaporated) at gas-liquid interfaces. This results in an update of all liquid structures to make the pressure correspond to the new volume.  The cluster pressure update algorithm returns the pressure in a single liquid cluster after its volume or configuration has changed. This algorithm is explained in detail in Appendix~\ref{sec:regula_falsi}. After the liquid content has changed, all instability criteria (\textbf{c1-c5}) (Sec.~\ref{sec:stabilitCriteria}) are checked. Before pore bodies can be filled (\textbf{c2-c4}), menisci must form, corresponding to criterion \textbf{c1}. Hence, first a list of bridges fulfilling \textbf{c1} is generated. Then new trimer units are formed by these bridges, eliminating all involved bridges from the list for further evaluation. Note that whenever a trimer is added to an already existing cluster due to \textbf{c1}, the entire cluster needs to be updated. The same holds if two clusters are merged by a trimer unit. When no more bridges fulfill criterion \textbf{c1}, we proceed with the other criteria in a similar fashion. This sequential procedure is chosen, since after fulfilling the higher numbered criterion, the preceding ones become even more unlikely and the liquid structure more stable. For criteria \textbf{c2-c4} also a list of unstable menisci is created, from which the most unstable configuration is chosen and processed (pressure update in the corresponding cluster). Then, the list of unstable menisci is updated. Note that alternative sequences of \textbf{c2-c4} were tested with similar outcome. When liquid is drained from the system, mainly criterion \textbf{c5} is relevant, however \textbf{c1-c4} are checked after the pressure update for consistency. 

Due to the instabilities, we obtain pressure differences in liquid structures that drive liquid transport through liquid films (see Sec.~\ref{sec:stabilitCriteria}, Eq.~\ref{eq:liqu_trans}). After the transport, the cluster pressure is recalculated to account for changes in volumes of liquid bodies. After the final pressure update, the time is incremented by $\Delta t$.
In Appendix we show the program sequence (Fig.~\ref{Fig:flowDiagram}) and list all simulation parameters (Tab.~\ref{tableSysPara}). 
\section{Changing liquid content on the pore scale} \label{sec:Validation}
Trimers are one of the building units for all larger clusters in a particle packing and the most common morphology after the liquid bridge for a wide range of saturation values \cite{ScheelPHD}. As they are well studied and rather simple, trimers are ideal validation and calibration clusters for the numerical model e.g. for trimer creation (criterion \textbf{c1}) and decay of single trimers (criterion \textbf{c5-1}). For the case of condensation, we study the distribution of cluster morphologies as a function of liquid content and compare to experiments by Scheel \textit{et al.} \cite{ScheelPHD}. We demonstrate the applicability of our method for the case of liquid injection at singular points until full saturation.
\subsection{Trimer formation and decay} \label{Sec:trimer}
\begin{figure}[htb]
\centering
\includegraphics[width=0.45\textwidth]{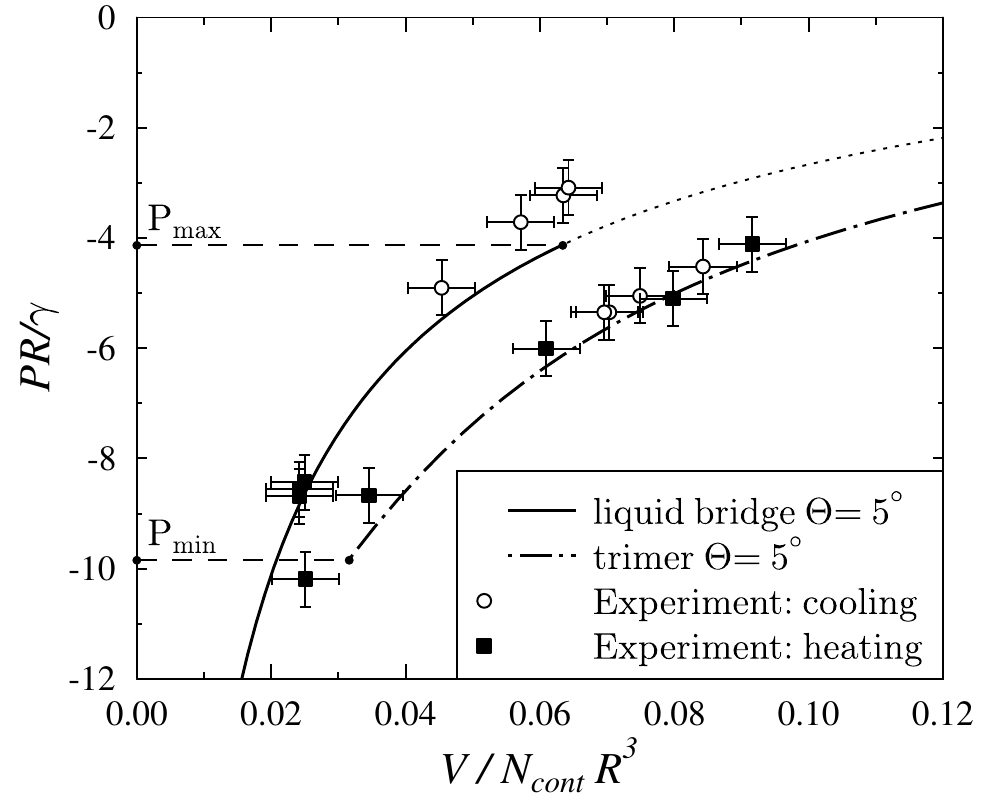} 
\caption{Dimensionless Laplace pressure of a trimer and liquid bridges as a function of their volumes normalized by the number of contacts $N_{cont}$ and the grain radius $R^3$. Trimers form at $P_{max}$ and decay at $P_{min}$ for a contact angle of $\theta=5^{\circ}$. The experimental data denoted by squares with error bars is taken from Ref.~\cite{ScheelPHD}. The lines show the simulation results.}
\label{TrimerPV}
\end{figure}

A trimer in our model consists of two menisci that are located on both sides of a single pore throat and three connected liquid bridges (Fig \ref{SphericalCap}b), analogous to observations with microtomography \cite{Scheel2008nature}. We calculate trimer creation by merging of two or three liquid bridges (criterion \textbf{c1}) and their decay (\textbf{c5-1}), and compare to experimental data  \cite{ScheelPHD} (see Fig.~\ref{TrimerPV}).

To reproduce the experiment we first simulate condensation into liquid structures. For this, a small amount of liquid $\Delta V_{cond} = 0.002R^{3}$ is added to each one of the three liquid bridges between the contacting grains in every time step of the simulation (see Fig. \ref{Fig:flowDiagram} in the Appendix). The liquid bridge pressure $P$ is updated to correspond to the changing volume. At a critical dimensionless value $P_{max}$, the bridges coalesce and a new trimer is created (Fig. \ref{TrimerPV}). At the same time the liquid pressure drops, since liquid is needed to fill the pore throat between the grains, thus decreasing the radius of the interface curvature. Then the process is inverted with liquid evaporating from the surface of the trimer. In this case, a small amount of liquid $\Delta V_{evap} = 0.01R^{3}$  is removed from the trimer in every time step while its pressure is adjusted. Eventually, the trimer decays into three liquid bridges at the critical minimal dimensionless pressure $P_{min}$. We confront our model prediction for this process with the experimental data in Fig.~\ref{TrimerPV}. To obtain good agreement the geometrical correction factor $\epsilon$ is set to 0.07 (see Eq.~\ref{eq:menVol}) and the drainage parameter $\kappa$ for the instability criterion \textbf{c5-1} (Eq.~\ref{equ:drainageInst}) was set to $\kappa =$0.15. 

\begin{figure}[htb]
\centering
\includegraphics[width=0.45\textwidth]{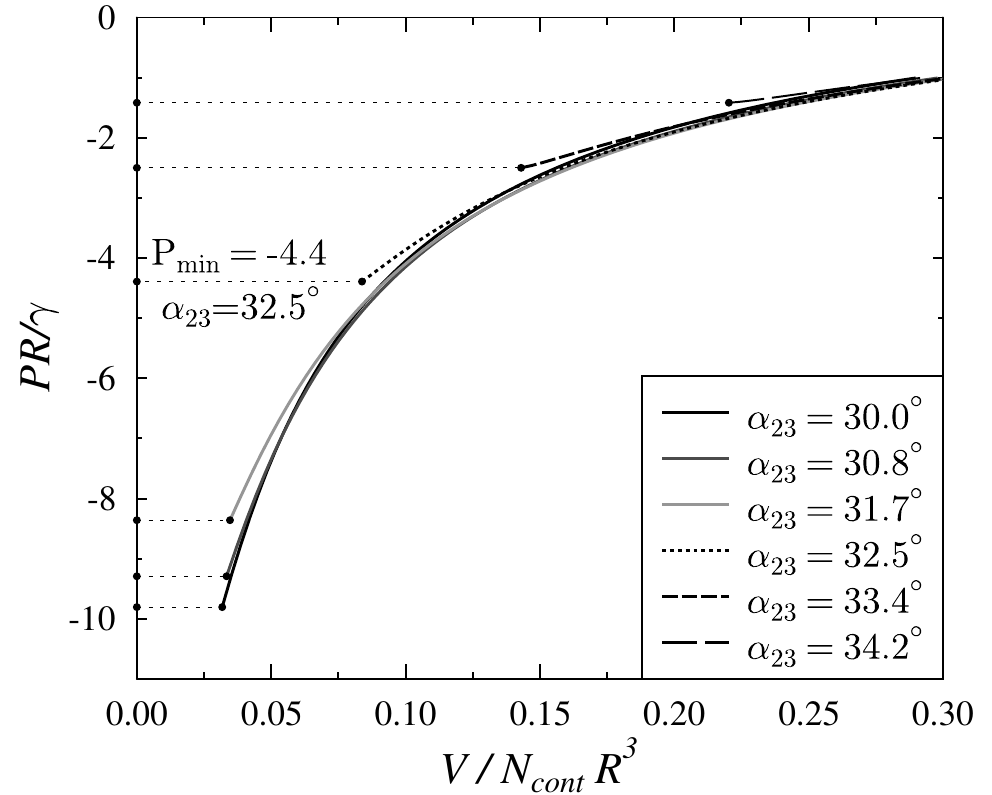}
\caption{Dimensionless Laplace pressure as a function of the normalized volume of a trimer for different opening angles $\alpha_{23}$ (see Fig.~\ref{Fig:cell}). Also indicated are the minimal Laplace pressures $P_{min}$ for which the trimer is still stable.}
\label{Fig:PressSepTrimer}
\end{figure}

The critical minimal pressure $P_{min}$ of a trimer strongly depends on the separation distance $S_{ij}$ between particles, or on the corresponding opening angles $\alpha_{ij}$ for a monodisperse particle packing. Since three liquid bridges are required for a trimer, we assume that a trimer can no longer exist if the separation distance $S_{ij}$ is larger than the rupture distance $S_c$ for one of the associated liquid bridges (see Eq.~\ref{equ:ruptureDist}). In Fig.~\ref{Fig:PressSepTrimer} we show the pressure-volume curves for a trimer in a particle configuration in which one of the opening angles $\alpha_{ij}=\alpha_{23}$ (see Fig.~\ref{SphericalCap}(b)) is increased. Note that $\alpha_{23}=30^\circ$ corresponds to the previously studied case of a closed contact (Fig.~\ref{TrimerPV}). As expected, large opening angles result in a high minimal pressure $P_{min}$, e.g. for $\alpha_{23} \geq 32.5^{\circ}$ no trimers can exist at a Laplace pressure $P_{trimer}<-4.4$. For imbibition this implies that trimers with particles not being in contact are generated at higher overall Laplace pressures compared to contacting ones.

\begin{figure}[htb]
\centering
\includegraphics[width=0.45\textwidth]{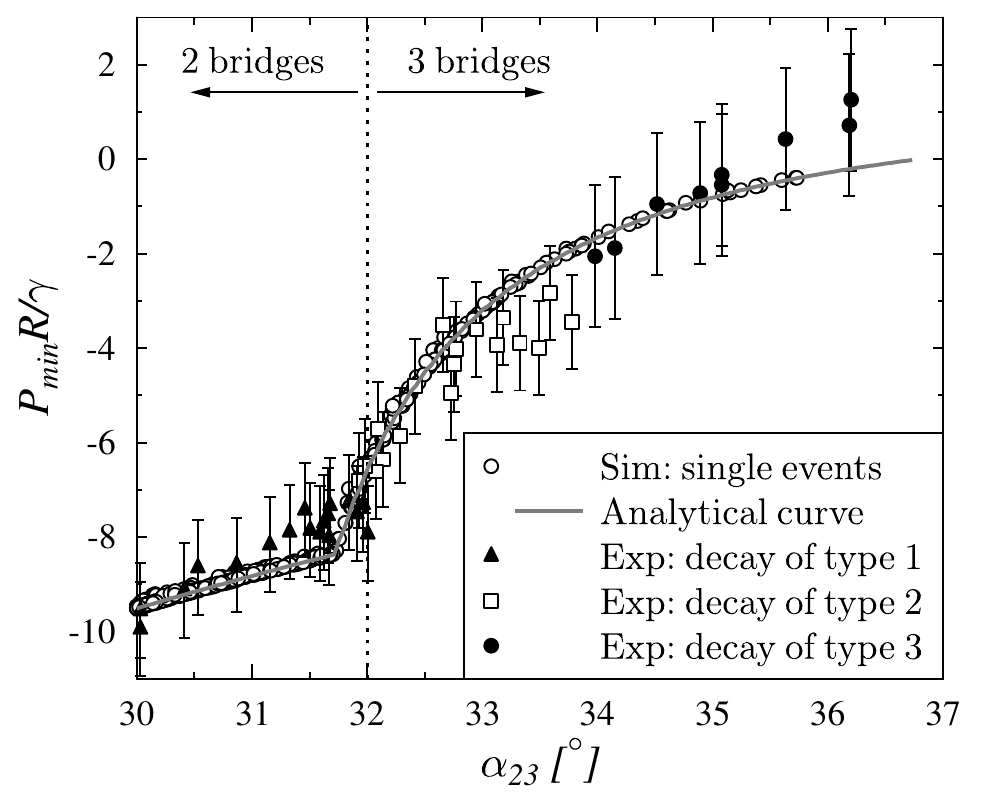}
\caption{Trimer decay: critical minimal Laplace pressure as a function of the opening angle $\alpha_{23}$ for trimers. Experimental data by Scheel \cite{ScheelPHD}. Three types of trimer decay are observed in the experiment: (1) decay into three liquid bridges, (2) decay into two liquid bridges, (3) slow transition into two liquid bridges. Only configurations with one gap opposite to the opening angle are shown.}
\label{Fig:trimerDecay}
\end{figure}


We demonstrate the validity of the above assumption by comparing the calculated dimensionless critical Laplace pressures $P_{min}$ as a function of the opening angle $\alpha_{23}$ for single trimers with measurements by Scheel \cite{ScheelPHD}.
In the experiment, trimers can decay into three ($30^\circ < \alpha_{23} < 32^{\circ}$) or two liquid bridges with fast ($32^\circ < \alpha_{23} < 34^{\circ}$) and slow ($\alpha_{23} > 34^{\circ}$) transition, as shown in Fig.~\ref{Fig:trimerDecay}. In the simulation liquid is evaporated from trimers in a dense packing through reduction of their volumes by $\Delta V_{evap}$ in every time step. In Fig.~\ref{Fig:trimerDecay} the critical pressure $P_{min}$ of the single trimer decays is shown for configurations with one gap opposite to the opening angle $\alpha_{23}$. Only decays of trimers are recorded in which the other two gaps are less than $10^{-2}R$. Our results are well within the error bars of the experimental measurements. We observe two different regimes for opening angles $\alpha_{23} > 31.7^\circ$ and $\alpha_{23} < 31.7^\circ$. The former is the result of the rupture of the liquid bridge, while the latter is due to the drainage criterion $H_{men}^{min}\geq\kappa R$ with $\kappa$ selected as $\kappa=0.15$ (see Eq.~\ref{equ:drainageInst}). The particular value of $\kappa$ was chosen to deliver the best agreement with experimental observations in Figs.~\ref{TrimerPV} and \ref{Fig:trimerDecay}. The critical pressure curve was also calculated analytically by considering a single trimer in the three particle configuration with a varying opening angle $\alpha_{23}$. This curve is also shown in Fig.~\ref{Fig:trimerDecay}. Small deviations of the simulated single trimer decays from the analytical solution are due to the finite size of $\Delta V_{evap}$ in the simulation.
\subsection{Morphogenetics of liquid clusters}\label{Sec:equil}
\begin{figure*}[htb]
\centering
\includegraphics[width=0.45\textwidth]{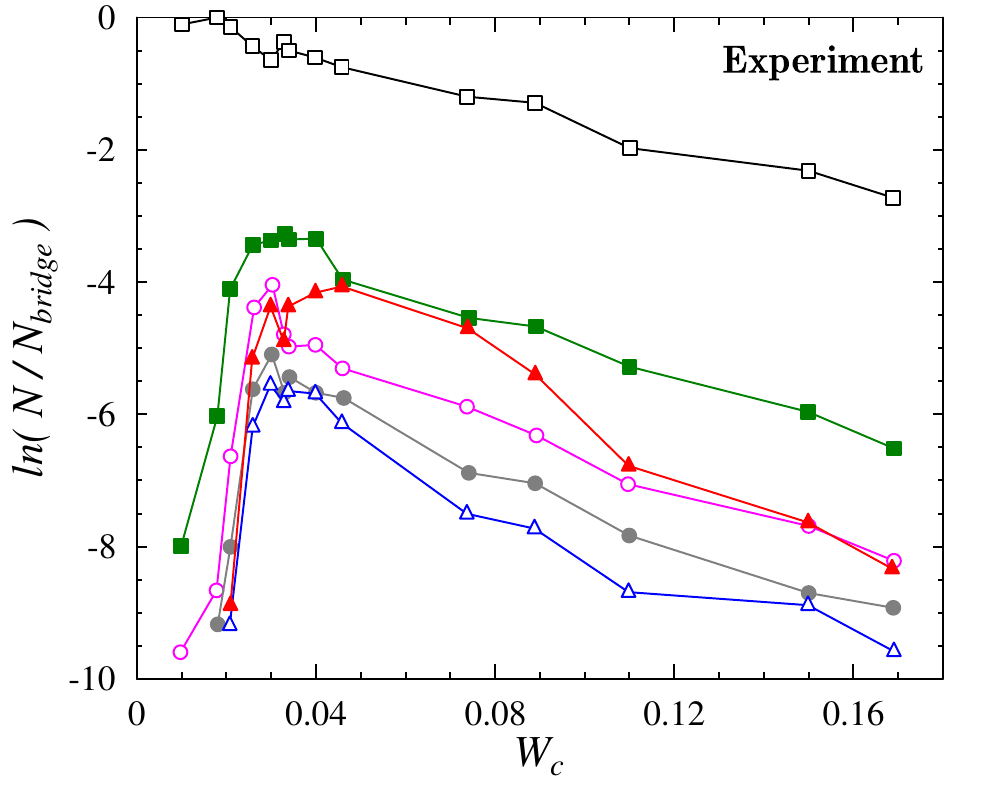}
\includegraphics[width=0.45\textwidth]{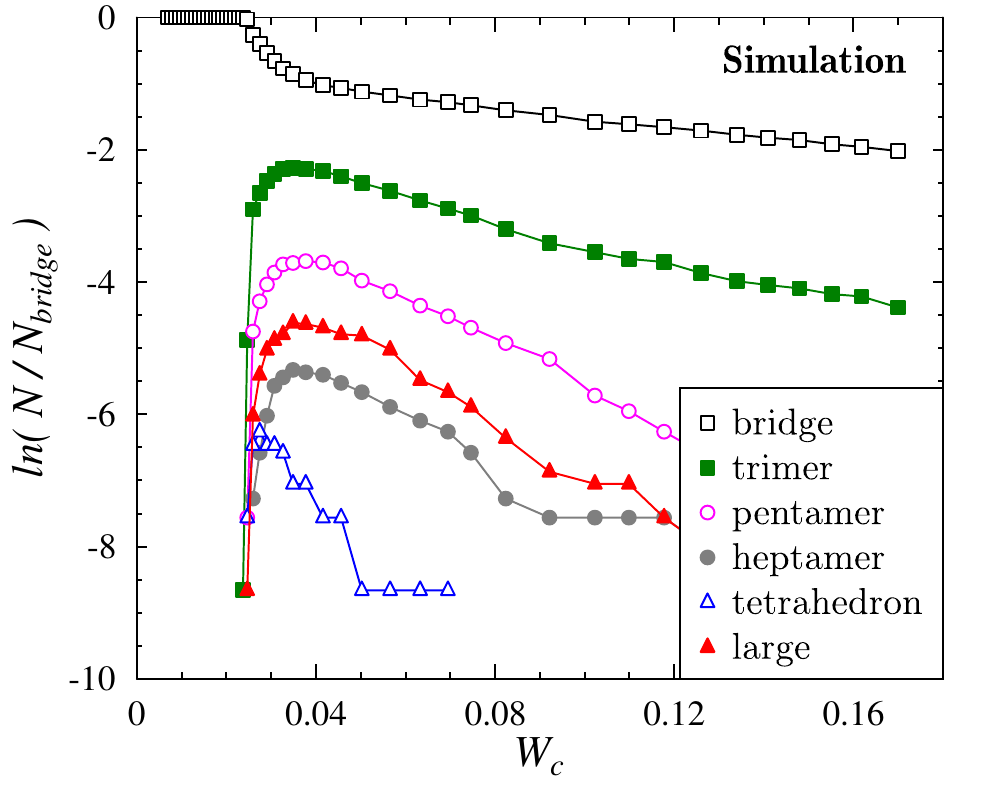}
\caption{(Color online) Number of liquid morphologies $N$ scaled with the maximal number of bridges $N_{bridge}$ in a packing with the density $\rho_{exp}=0.57\pm 0.01$ (experiment) respectively $\rho_{sim}=0.57$ (simulation) as a function of the liquid content $W_{c}$. Experimental data from Ref.~\cite{ScheelPHD}.}
\label{Fig:Nstruct}
\end{figure*}

\begin{figure}
\centering
\includegraphics[width=0.45\textwidth]{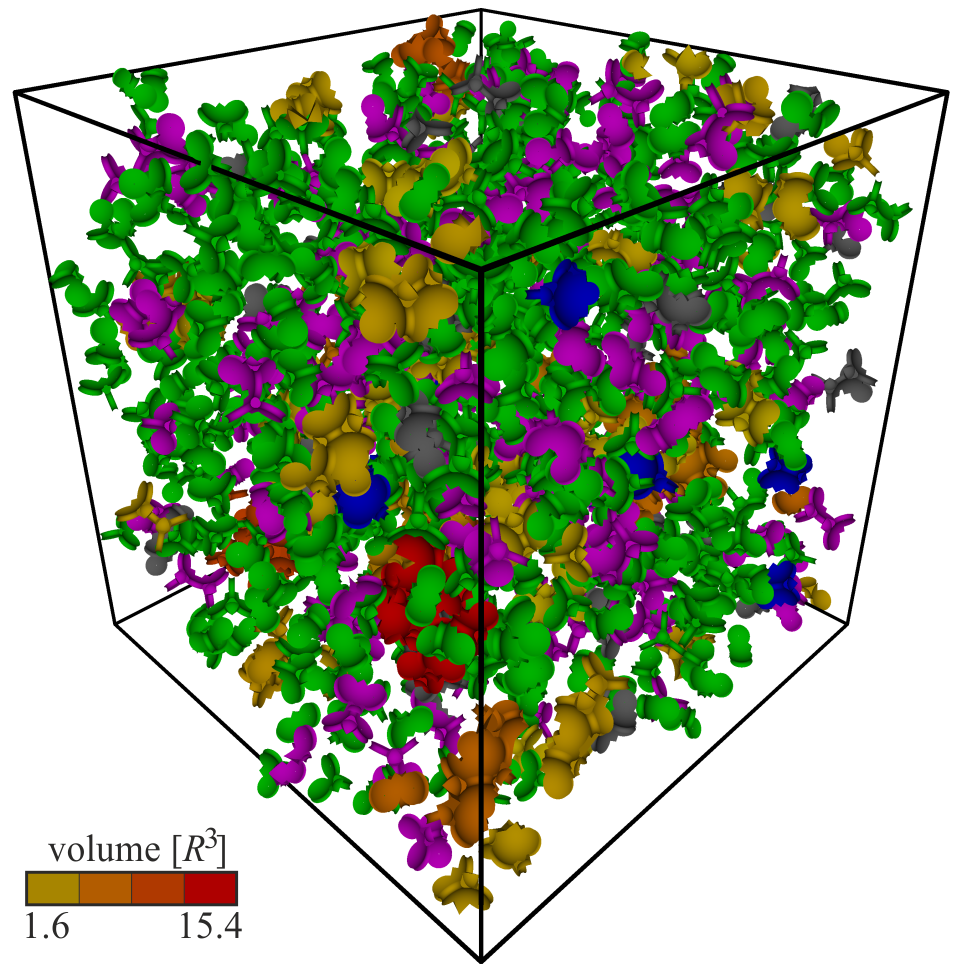}
\caption{(Color online) Snapshot of the sample with different cluster morphologies at $W_{c}=0.03$. Only liquid bodies of the clusters are shown. Single liquid bridges are omitted for clarity. Green: trimers, magenta: pentamers, grey: heptamers, blue: tetrahedral clusters. Large clusters are colored according to their volume with a color map ranging from yellow to red.}
\label{Fig:SnapshotClusters}
\end{figure}  
Studies on single liquid clusters are important for the model validation, however liquid in a random packing leads to a multitude of clusters of different topology and size. Their statistic was evaluated from microtomographic images for a random close packing with density $\rho = 0.57 \pm 0.01$ at different liquid contents $W_c$ \cite{ScheelPHD}. 
In the simulation we increase $W_{c}$ by condensing
liquid into clusters and single liquid bridges starting with a pendular regime. A small amount of liquid proportional to the surface is added to every structure in each time step. The obtained distributions are shown and compared with experimental ones in Fig.~\ref{Fig:Nstruct}. A snapshot of the sample with different cluster morphologies at $W_{c}=0.03$ is shown in Fig.~\ref{Fig:SnapshotClusters}. At low liquid contents only liquid bridges exist, while larger clusters start to emerge later. Above the pendular state, the number of clusters rapidly increases and reaches a maximum which depends on the corresponding cluster size. After the maximum of the respective morphology is reached, the clusters predominantly merge and build larger structures. Note that finally only one single percolating cluster remains. Our simulation reproduces the general trend observed in the experiment.
However, some deviations are present (in particular for tetrahedral clusters). One of the main reasons for these deviations, apart from the approximate nature of instability criteria \textbf{c1-c5}, is a finite sample size of the simulation with only up to 11 single tetrahedral clusters being observed (see Fig.~\ref{Fig:SnapshotClusters}).
\subsection{Haines Jumps in Single Cluster Growth}\label{Sec:Haines}
\begin{figure*}[htb]
\centering
\includegraphics[width=0.8\textwidth]{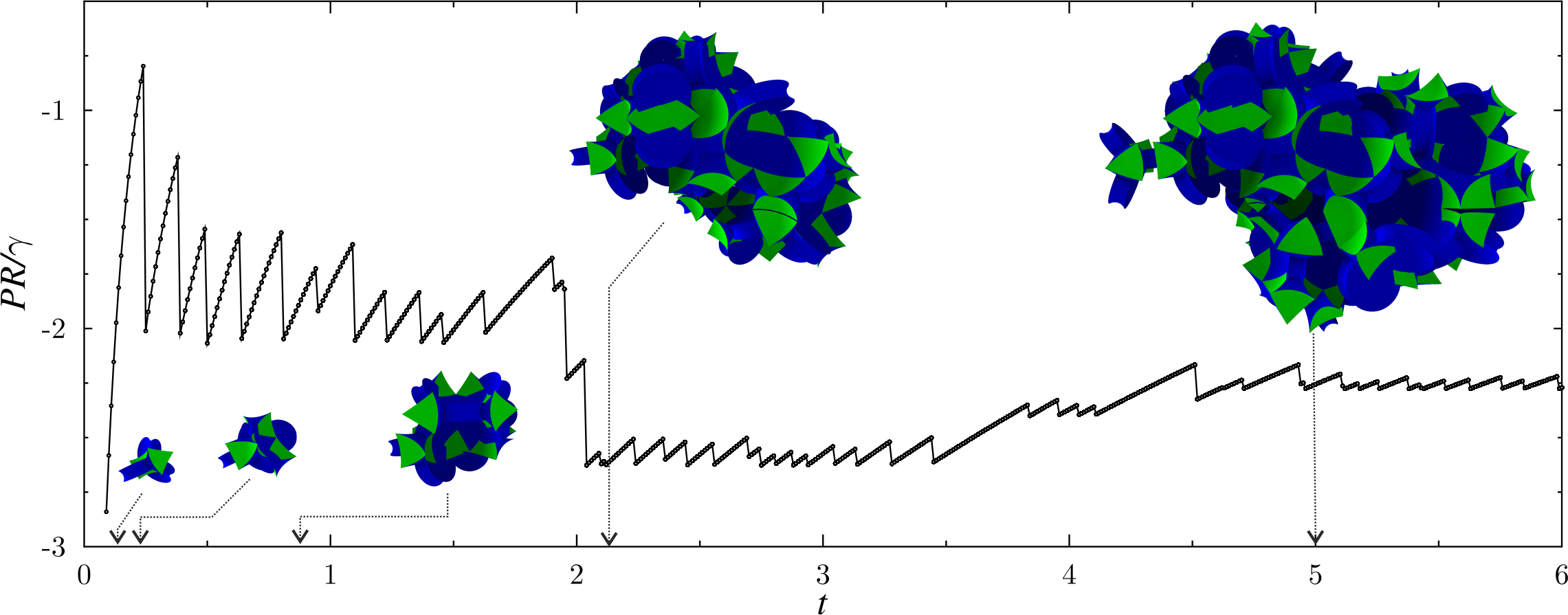}
\caption{(Color online) Dimensionless Laplace pressure of a liquid cluster as a function of time, when liquid is injected at a constant flow rate at a fixed point with evolving liquid cluster in terms of its liquid body (insets).}
\label{Fig:clusterSnap}
\end{figure*}
When liquid is injected into the particle packing at a singular point, pressure drops in the single liquid body of the emerging cluster occur. Pressure drops originate from rapid imbibitions of single pore bodies, so called Haines jumps, and have been observed in experiments (e.g. Ref.~\cite{Berg2013}). Pressure drops are due to volume conservation with respect to the new configuration, resulting in increasing curvature of menisci. Starting from a liquid content in the pendular regime ($W_{c} \approx 1\%$), we inject liquid with a constant flux into the center of the sample. Initially, liquid bridges at the injection point grow and eventually coalesce to a trimer that grows until the first instability occurs. Then the cell pore body is imbibed, creating a tetrahedral cluster. In the following, an increasing number of pore bodies is imbibed, resulting in a drop of Laplace pressure inside the cluster (see Fig.~\ref{Fig:clusterSnap}). In the beginning, the dimensionless cluster pressure increases and reaches the value of $-0.79$, followed by a sharp drop. Each drop is associated with the local instability of filling a new pore body or throat. In Fig.~\ref{Fig:clusterSnap} we observe decreasing  magnitudes of pressure drops with increasing cluster size. This can be explained by the fact that the amount of liquid needed to fill the next pore body in relation to the total cluster volume decreases with increasing cluster volume. Note that the pressure converges to a approximately constant value. This value is given by the minimal possible pressure of a "critical" liquid bridge of the cluster which has a maximal separation distance between the particles (see Fig.~\ref{Fig:PressSepTrimer}). The morphology of the growing cluster at different time steps is shown in Fig.~\ref{Fig:clusterSnap}. Prior to the first pressure drop, the liquid cluster is a trimer ($t=0.15$), then a tetrahedral cluster ($t=0.25$), and later on it grows in all directions in a compact manner, even with single trimer units being connected to it ($t=5.01$). The compact shape of the cluster is due to the pressure equilibration time scale which is set by the conductance coefficient $\mu$. For $\mu=0.01$ the liquid transport through surface films is slow compared to the cluster growth that is triggered by the liquid injection rate. Therefore the cluster grows in a compact manner mainly through imbibition of pore bodies. The pressure-volume relationship for a trimer like the one at the beginning of cluster growth (Fig.~\ref{Fig:clusterSnap}) has been compared with experimental results in Fig.~\ref{TrimerPV}. For larger clusters no experimental data is available. However, the convergence to a constant pressure level during the cluster growth can be linked to experimental observations which state that the pressure in equilibrated liquid morphologies remains constant in a wide range of saturation levels \cite{Scheel2008nature}.

\section{Summary and Conclusions} \label{sec:Conclusion}
We propose a model for fluid saturation in random packings that can cope with arbitrary liquid contents ranging from dry to full saturation with good accuracy. Volume is used as a control variable and pressure is calculated based on the volume conservation for every liquid cluster contributing to the liquid content in the porous sample. This innovation is of particular importance since the volume respectively the water content is much easier to access experimentally than the pressure. The model is able to reproduce all kinds of liquid clusters on the grain scale observed in experiments like bridges, trimers, pentamers and so on. This was shown on a number of experimental benchmark problems.
We demonstrated that by using simple geometrical approximations, fairly good agreement with experiments can be obtained at only a small portion of effort required to model the true shape of the interfaces by for example energy minimization methods \cite{Brakke1996}. Since we use volume as a control variable, we are capable of calculating pressure drops in single clusters due to Haines jumps, as shown in the example of Sec.~\ref{Sec:Haines}. The approach is not limited to static pore spaces, and can be used with minor modifications for the study of unsaturated deformable granular packings. Studies on the interaction of the pore space with the pressure field are of particular interest for questions of soil behavior. 
\begin{acknowledgments}
We would like to thank Ciro Semprebon and Martin Brinkmann for providing tabulated values characterizing single liquid bridges.
The research leading to these results has received funding from the People Programme (Marie Curie Actions) of the European Union's Seventh Framework Programme FP7 under the MUMOLADE ITN project (Multiscale Modelling of Landslides and Debris Flow) with REA grant agreement n$^{\circ}$ 289911, as well as from the European Research Council Advanced Grant no.~319968-FlowCCS and the DFG under PiKo SPP 1486 HE 2732/11-3.
\end{acknowledgments}
\appendix 
\section{Simulation parameters}

Simulation parameters are summarized in Tab.~\ref{tableSysPara}.

\begin{table}
\caption{\label{tableSysPara} Summary of simulation parameters.}
\begin{ruledtabular}
\begin{tabular}{lll}
 Label & Value & Definition  \\ \hline 
    $N_p$ & 2000 & number of particles\\
    $\rho$ & 0.57 & packing density\\
    $R$ & 1.31 & particle radius \\
    $\Delta t$ & 0.001 - 0.01 & time step\\
    $\Theta$& $5^{\circ}$ & contact angle \\
    $\gamma$& 1 & surface tension\\
    $\mu$ & 0.01 & conductance coefficient\\
    $\epsilon$& 0.07 & geometrical correction parameter\\
    $\kappa$ & 0.15 & drainage parameter for meniscus\\     
    \end{tabular}{}
\end{ruledtabular}
\end{table}

\section{Pressure update algorithm}\label{sec:regula_falsi}
The main aim of the pressure update algorithm is to find the new cluster radius $R_{men}$ such that the $V_{c}(R_{men})-V_{old} = 0$ is satisfied, where $V_{old}$ denotes the cluster volume before and $V_{c}$ after the pressure update.
\begin{figure}[htb]
\centering
\includegraphics[width=0.45\textwidth]{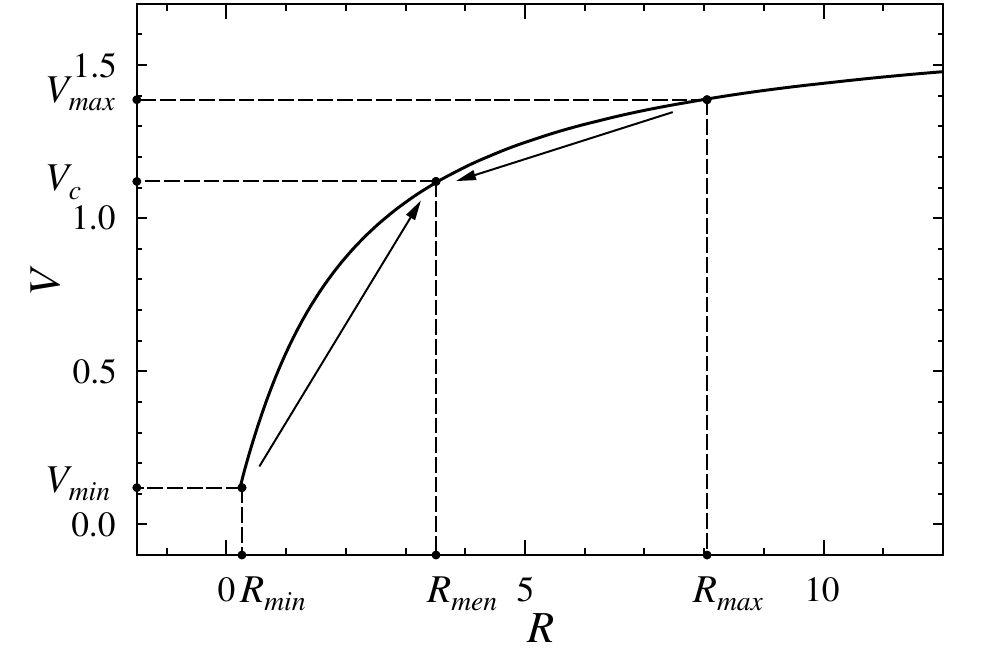}
\caption{Calculation of the final meniscus radius $R_{men}$ which lies between the radius  with the minimal volume $V_{min}$ and the one with a larger volume $V_{max}$ here shown for a trimer.}
\label{Fig:clusterAlgorithm}
\end{figure}
\begin{figure}[htb]
\centering
\includegraphics[width=0.4\textwidth]{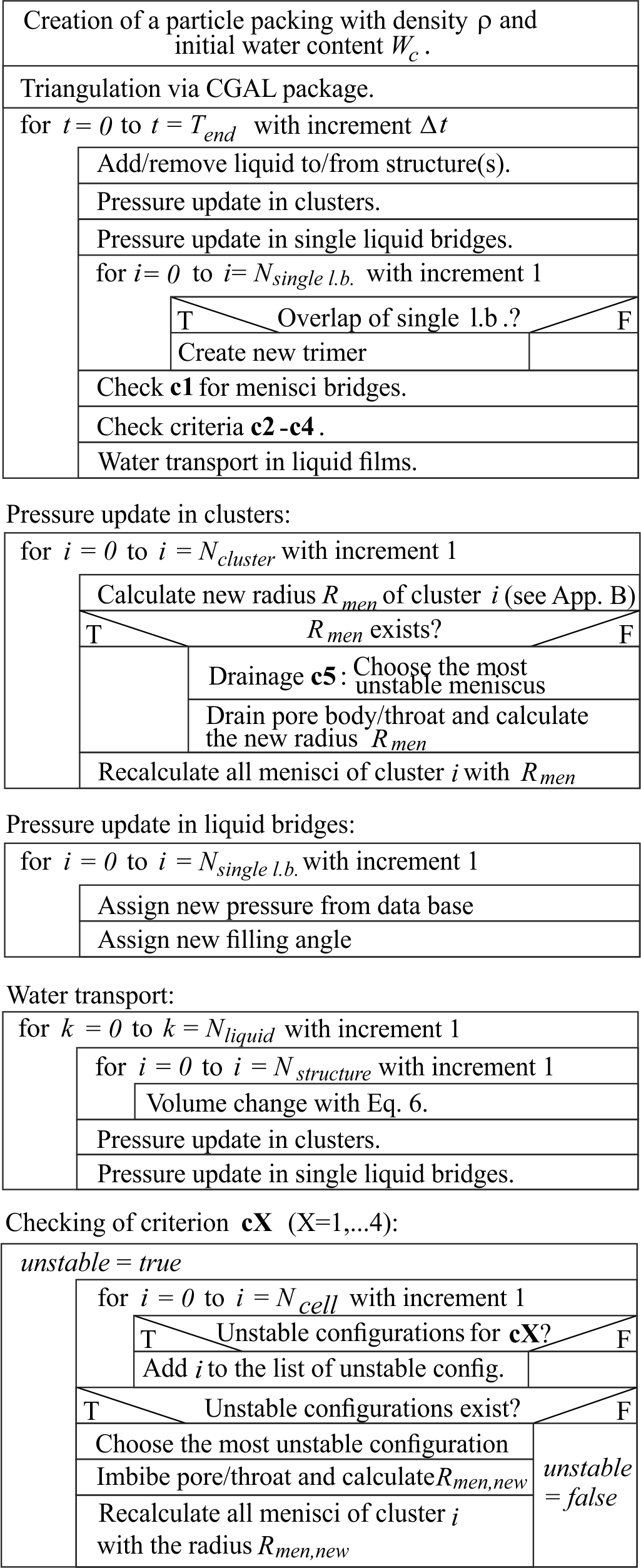}
\caption{Scheme of the simulation progress.}
\label{Fig:flowDiagram}
\end{figure} 
This root-finding problem is numerically solved using the false position method (Regula Falsi) see Fig.~\ref{Fig:clusterAlgorithm}. The target radius of the recalculated cluster menisci radius $R_{men}$ which corresponds to the volume $V_{c}$ is located between $R_{min}$ and $R_{max}$. $V_{min}$ and $V_{max}$ denote the cluster volumes calculated with the radius $R_{min}$ respectively $R_{max}$. The value $R_{min}$ is the minimal radius for which the recalculated cluster is still stable. This lower threshold is given by the pressure of the most unstable meniscus taking into account the associated liquid bridges. An implication of choosing a lower threshold $R_{min}$ is that in some cases the available volume $V_{old}$ is smaller than the minimal volume of the new (recalculated) cluster $V_{min}$: $V_{old}<V_{min}(R_{min})$. This is for example the case when a new pore body should be imbibed by liquid from a small cluster which has not enough liquid to fill the pore body and create a stable cluster interface. Since in this case the imbibition event cannot occur, the corresponding instability is saved in the list of impossible instabilities which are not tested again in the current time step. The maximal radius $R_{max}$ can be chosen depending on the cluster volume before the volume change or instability event. It should be noted here that there is a saturation volume for every cluster which corresponds to the case where all menisci of the cluster have infinite radius. In real simulations this volume is never reached because an instability will appear before. Therefore it is always possible to find a proper value for $R_{max}$  such that $V_{c}<V_{max}$. For decreasing cluster volumes, the cluster radius before the pressure update is used as $R_{max}$. If the cluster volume increases, $R_{max}$ is set to a multiple of the cluster radius before the pressure update. In the end, the pressure update algorithm delivers the radius of the menisci in the cluster $R_{men}$ for which the volume is conserved.
\bibliographystyle{ieeetr}
\bibliography{paper.bbl}
\end{document}